\documentclass[3p,12pt,authoryear]{elsarticle}
\usepackage[utf8]{inputenc}
\newcommand\grl{Geophys.~Res.~Lett.}%
\newcommand\icarus{Icarus}%
\newcommand\planss{Planet.~Space~Sci.}%
\usepackage{tabto}

\newenvironment{tabs}[1]
 {\flushleft\TabPositions{#1}}
 {\endflushleft}
 
\begin{document}

\huge
\begin{center}
    The Saturn Ring Skimmer Mission Concept:
    \Large{The next step to explore Saturn's \\ rings, atmosphere, interior, and inner magnetosphere}

\vspace{1cm}

\normalsize{}
   A white paper submitted to the 2023 Planetary Science Decadal Survey\\
\vspace{0.4cm}
August 15, 2020%\today

\vspace{1cm}

\normalsize
\noindent Matthew~S.~Tiscareno$^1$, Mar~Vaquero$^2$, Matthew~M.~Hedman$^3$, Hao~Cao$^4$, Paul~R.~Estrada$^5$, Andrew~P.~Ingersoll$^6$, Kelly~E.~Miller$^7$, Marzia~Parisi$^2$, David.~H.~Atkinson$^2$, Shawn~M.~Brooks$^2$, Jeffrey~N.~Cuzzi$^5$, James~Fuller$^6$, Amanda~R.~Hendrix$^8$, Robert~E.~Johnson$^9$, Tommi~Koskinen$^{10}$, William~S.~Kurth$^{11}$, Jonathan~I.~Lunine$^{12}$, Philip~D.~Nicholson$^{12}$, Carol~S.~Paty$^{13}$, Rebecca~Schindhelm$^{14}$, Mark~R.~Showalter$^1$, Linda~J.~Spilker$^2$, Nathan~J.~Strange$^2$, Wendy~Tseng$^{15}$

\vspace{0.5cm}
\footnotesize
$^1$SETI~Institute, $^2$NASA/Caltech~Jet~Propulsion~Laboratory, $^3$University~of~Idaho, $^4$Harvard~University, $^5$NASA~Ames~Research Center, $^6$Caltech, $^7$Southwest~Research~Institute, $^8$Planetary~Science~Institute, $^9$University~of~Virginia, $^{10}$University~of~Arizona, $^{11}$University~of~Iowa, $^{12}$Cornell~University, $^{13}$University~of~Oregon, $^{14}$Ball~Aerospace, $^{15}$National~Taiwan~Normal~University

\vspace{1cm}
\normalsize
At time of submission this paper had over 60 co-signers, in addition to the 24 authors. 
\\A list of supporters can be found at \href{https://www.seti.org/RingSkimmerWP20_supporters}{www.seti.org/RingSkimmerWP20\_supporters}\\
\vspace{0.7cm}
\resizebox{6.5in}{!}{\includegraphics{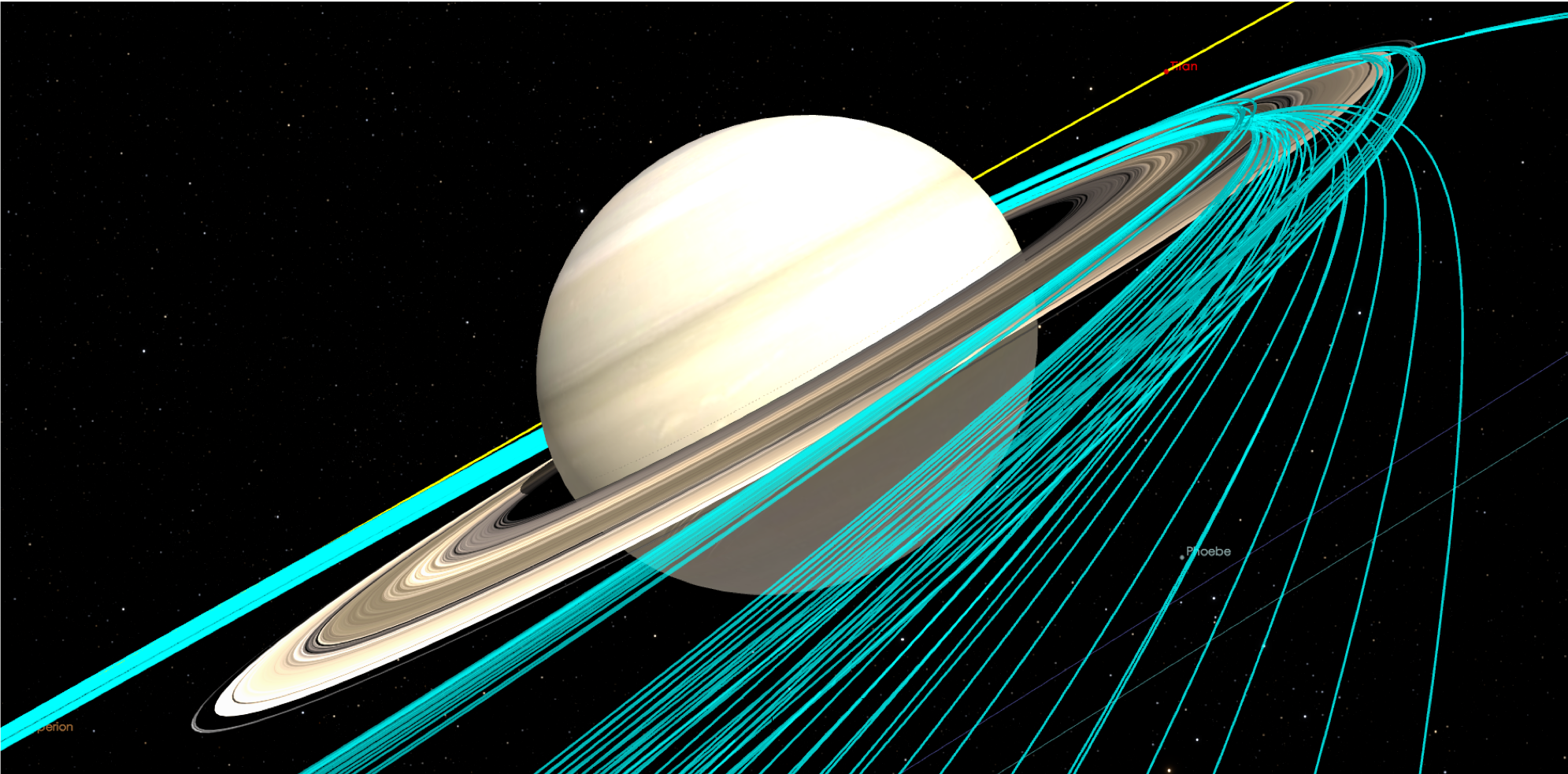}}
\end{center}

\thispagestyle{empty}
\newpage
\pagenumbering{arabic}

\normalsize
 {\bf Summary: {\em The innovative Saturn Ring Skimmer mission concept enables a wide range of investigations that address fundamental questions about Saturn and its rings, as well as giant planets and astrophysical disk systems in general.}} This mission would provide new insights into the dynamical processes that operate in astrophysical disk systems by observing individual particles in Saturn's rings for the first time. The Ring Skimmer would also constrain the origin, history, and fate of Saturn's rings by determining their compositional evolution and material transport rates.  In addition, the Ring Skimmer would reveal how the rings, magnetosphere, and planet operate as an interconnected system by making direct measurements of the ring's atmosphere, Saturn's inner magnetosphere and the material flowing from the rings into the planet. At the same time, this mission would clarify the dynamical processes operating in the planet's visible atmosphere and deep interior by making extensive high-resolution observations of cloud features and repeated measurements of the planet's extremely dynamic gravitational field. {\bf{\em Given the scientific potential of this basic mission concept, we advocate that it be studied in depth as a potential option for the New Frontiers program.}}

\vspace{0.2cm}
\textbf{Background and Motivation:}  The enormously successful Cassini mission revolutionized our understanding of the entire Saturn system and  substantially addressed dozens of scientific questions and mysteries.  As always happens with ground-breaking science, the data provided by Cassini has raised important new questions about Saturn and its rings, including:

\vspace{0.1cm}
\textbf{{\em 1. How do rings and particle-rich disks work at the particle level?}} Saturn's rings span 300,000 km, but are composed primarily of particles millimeters to meters across \citep{PRSBook}. Cassini documented a variety of structures in the rings that reflect differences in how these particles aggregate, fragment, and interact in different environments  \citep{CuzziChapter18, Rgogfrings19}. Furthermore, Cassini found evidence that larger objects (essentially ``proto-moons") were orbiting within the rings and interacting with the surrounding ring material \citep{Spahnchapter18}.  However, Cassini never had the resolution to directly image individual ring particles, and so fundamental aspects of the physical properties and impact dynamics of ring particles are still not well constrained \citep{Salochapter18}. The particle-level processes operating in Saturn's rings should be analogous to those occurring in late-stage protoplanetary disks like the one that gave rise to our solar system, so more detailed information about these phenomena would allow us to better understand not only Saturn's rings, but also our own origins. 

\vspace{0.1cm}    
\textbf{{\em 2. What are the origin, history, and fate of Saturn's rings?}} Measurements made during Cassini's Grand Finale revealed surprisingly high fluxes of ring material into Saturn \citep{Waite18, Mitchell2018,Hsu18, Perry18, Odonoghue19}. A steady-state interpretation of these fluxes, along with the low mass of the rings \citep{Iess19}, suggests that Saturn's rings could be young and/or short-lived. However, Cassini could only briefly sample this material, and its temporal variations and spatial distribution are not well constrained. Furthermore, the equatorial mass flux has a surprisingly low water fraction \citep{Waite18}, which may imply that non-water constituents are preferentially removed from rings, further complicating the situation \citep{Crida19}. Also, extrinsic bombardment and the resulting ballistic transport of ejecta likely have a profound effect on ring structure \citep{EstradaChapter18}, but the evolution rates are not well constrained. Therefore, much remains uncertain about the history and future of both the rings and the Saturn system as a whole \citep{Crida19, Estrada19}.

\vspace{0.1cm}
\textbf{{\em 3. How do the rings, planet, and magnetosphere influence each other's structure and evolution?}} 
Saturn is the only planet in our Solar System that is surrounded by a broad, dense ring system and that has a magnetic field that is nearly perfectly aligned with the planet's spin axis. This creates a unique interconnected system, with material from the rings flowing through the magnetosphere into the upper atmosphere \citep{Perry18, Odonoghue19}, the rings themselves influencing spatial distribution of plasma and neutrals \citep{Waite05, Farrell2018, Sulaiman2018, Dougherty2018, Khurana2018}, and asymmetries in the magnetosphere affecting the spatial and temporal distribution of dust-sized particles around the rings \citep{Mitchell13, Chancia19, Cao2020}. Cassini's trajectory did not allow it to fully document the distribution of neutrals, plasmas, and currents between the rings and the planet, or the asymmetries and time variations in these quantities. Cassini also found that the ring system has its own atmosphere, which is the solar system's only significant disk-shaped atmosphere. However, its complete composition, chemistry, and temporal variability are primarily inferred from measurements outside the rings \citep{Tokar05,Waite05,Tseng13,Miller20}. Moreover, although Cassini documented the seasonality and kinematics of the ghostly spokes, their origins and basic mechanism remain mysterious \citep{Mitchell13}. This information is needed not only to clarify the history and evolution of the inner Saturn system, but also to make sensible predictions about the atmospheric and magnetospheric properties of exoplanets whose evolution may be influenced by ring systems or other disks \citep{SC11}. 

\vspace{0.1cm}
\textbf{\em 4. What are the dominant dynamical processes in Saturn's atmosphere?}  Close-up images of Saturn's atmosphere taken during Cassini's Grand Finale revealed previously unexpected atmospheric structures at scales of $1-10$~km \citep{Ingersoll18}.  Notable features include puffy clouds resembling terrestrial cumulus, shadows indicating cloud height, dome- and bowl-shaped cloud structures indicating upwelling and downwelling, and thread-like cloud filaments that remain coherent over distances of 20,000~km. Cassini observations of these phenomena  are extremely limited, and more data about their spatial and temporal structure would provide important insights into the fundamental physical processes operating in giant planet atmospheres, including condensation, convection, and precipitation.

\vspace{0.2cm}
\textbf{{\em 5. What is happening deep inside Saturn?}} Cassini's Grand Finale revealed that Saturn's gravitational field is far more dynamic or complex than anyone had expected. Variations in the gravitational accelerations felt by Cassini during different passages between the planet and the rings imply that Saturn has an array of asymmetries in its deep interior that are unlike anything seen at Jupiter, and it is not even known whether these vary with time and/or azimuth \citep{Iess19}.  Saturn's rings also contain a variety of structures that are generated by oscillations and asymmetries inside the planet \citep{MarleyPorco93,  HN13, HN14, French19, Hedman19}. These dynamical phenomena can provide important insights into the internal structure of both Saturn in particular and giant planets in general \citep{Marley14, Fuller14, Mankovich19}.

\vspace{0.1cm}
%\pagebreak

The previous Decadal Survey identified high-level, long-term science questions that align with the questions we've articulated.  Specifically, the \textit{Building New Worlds} theme asks ``How did the giant planets and their satellite systems accrete, and is there evidence that they migrated to new orbital positions?'' and the \textit{Workings of Solar Systems} theme asks ``How do the giant planets serve as laboratories to understand Earth, the solar system, and extrasolar planetary systems?'' and ``How have the myriad chemical and physical processes that shaped the solar system operated, interacted, and evolved over time?''

During the previous Planetary Science Decadal Survey, just two missions to Saturn or its rings were studied in detail: a dedicated ring observer and an atmospheric entry probe \citep{Beebe10,Nicholson10}. Only the latter was found a viable candidate for the New Frontiers program. However, \citet{Vaquero19} recently introduced an innovative ``Ring-Skimming'' trajectory that  enables an entirely new type of mission to the inner Saturn system.

\begin{figure}[t!]
\begin{center}
%\vspace{-.7cm}
\includegraphics[width=0.3\textwidth,viewport=6.5cm 8cm 14cm 14.4cm,clip]{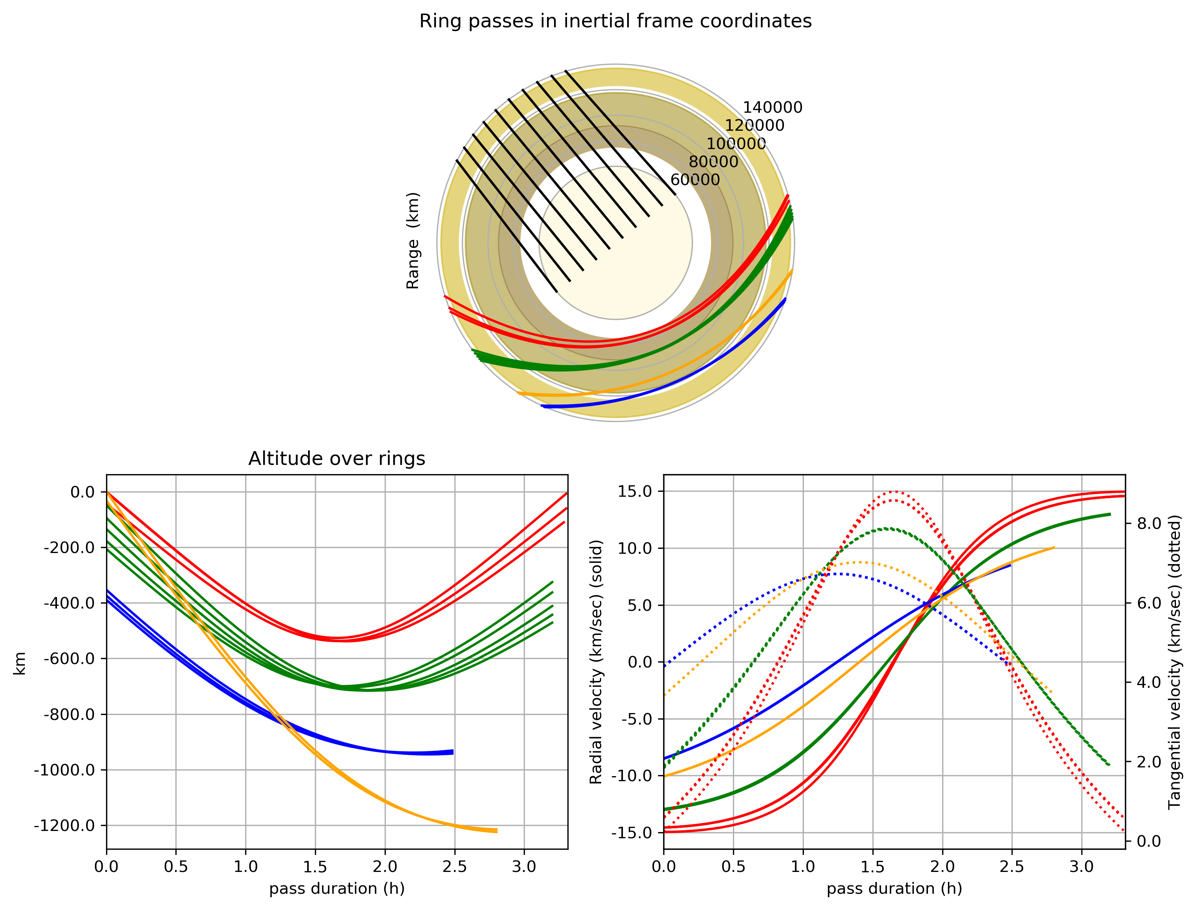}
\includegraphics[width=0.65\textwidth,viewport=0cm 0cm 20.5cm 8cm,clip]{tour1}
\caption{\footnotesize{Polar plot (left) illustrating 13 passes over Saturn's rings corresponding to the 162-day long prototype ballistic tour; the altitude (middle) and relative velocity (right) curves represent the passes over the rings. The black solid lines on the left panel represent the region of the rings shadowed by the Sun and, thus, in eclipse. This ring-skimming trajectory is ballistic and exploits four Titan gravity assists. For reference, the ring passes are color coded and grouped by Titan flybys.  Figure from \citet{Vaquero19}.
\label{VaqueroFig}}}
\end{center}
\end{figure}

\vspace{0.2cm}
\textbf{The Ring Skimmer Mission Concept:} \citet{Vaquero19} presented a prototype ballistic tour that passes repeatedly at low altitude over Saturn's main rings (see Figure~\ref{VaqueroFig}). In the span of only 162 days, and without using any propellant, this prototype tour covers the main ring regions in 13 low-altitude passes, but nothing prevents a much longer ring-skimming tour from being designed.  Flybys occur on the sunlit side of the rings, but skimming over the unlit side could be accomplished with a tweak in the Titan flyby design or the addition of a maneuver. In order to protect the spacecraft from particle impact, the orbit node crossings are located outward of the F~ring, which limits the periapse distance to $\sim$74,000~km (the inner edge of the C~ring). 

A spacecraft following this trajectory would get roughly 100 times closer to the rings than Cassini was when its best ring images were taken, and so would be able to obtain images with far better spatial resolution. In addition, the spacecraft's proximity to the rings allows for in-situ measurements of the material surrounding the rings, including the ring's tenuous atmosphere, dust released by collisions into the ring particles or levitated by electromagnetic forces, and material flowing along magnetic field lines from the rings into the planet. Furthermore, the spacecraft will be close enough to the rings for it to feel their gravity, enabling the mass density to be mapped at higher fidelity than previously possible. At the same time, the spacecraft gets close to Saturn itself many times, allowing for repeated measurements of the planet's asymmetric gravitational field, high-resolution images of its atmospheric clouds, and in-situ observations of the inner magnetosphere. Such a mission could therefore potentially address all five of the above science questions using current technology and with feasible costs.

In order to better evaluate the scientific potential of this basic mission concept for addressing the above science questions, we have developed the following science goals and specific measurement objectives that could be met by a spacecraft on a Ring-Skimming trajectory. Each of the five goals is directly linked to one of the above science questions, while the objectives could be met with the following primary investigations: 
\vspace{-.3cm}
\begin{tabs}{2.3in,4.6in}
IMG = (Color) Imaging	\tab MS = Mass Spectrometer \tab PW = Plasma Wave \\
GS = Gravity Science\tab DD = Dust Detector \tab MAG = Magnetometer \\
\end{tabs}

%\vspace{0.3cm}
\textbf{\em Science Goal \#1: Characterize the processes that shape particle-rich astrophysical disks.}  Particles in Saturn's rings are constantly interacting with each other through their own gravity and collisions, generating a diverse array of temporary agglomerations of various sizes, shapes, and orientations. At the same time, the dynamics of the collisions among ring particles influence how large particles can grow within different parts of the ring. The basic physical processes of gravity, collisions, aggregation, and fragmentation operating in these different ring regions should be analogous to those occurring in any particle-rich disk, and so the rings provide a natural laboratory for investigating these phenomena. We can determine the relative importance of these processes under various conditions by relating  the physical and dynamical properties of individual ring particles to the larger scale ring structures at various locations. We can also ascertain how objects accrete, move, and break apart within a disk by observing interactions between typical ring particles and larger objects. These considerations led us to identify the following measurement objectives:
\begin{itemize}
\vspace{-0.2cm}
\item \textbf{Objective 1-1:} {\em Measure the properties (size, shape, color) of individual ring particles at various locations in Saturn's rings.} (IMG)
\vspace{-0.3cm}
\item \textbf{Objective 1-2:} {\em Determine collision parameters and outcomes at various locations in Saturn's rings.} (IMG)
\vspace{-0.3cm}
\item \textbf{Objective 1-3:} {\em Measure clumping at various scales in different regions of Saturn's rings.} (IMG)
\vspace{-0.3cm}
\item \textbf{Objective 1-4:} {\em Observe propeller objects caused by embedded objects within the rings and their interactions with surrounding ring material.} (IMG)
\vspace{-0.3cm}
\item \textbf{Objective 1-5:} {\em Characterize ring-moon shapes and surfaces at close range.} (IMG)
\vspace{-0.3cm}
\item \textbf{Objective 1-6:} {\em Measure the radial mass distribution of Saturn's rings.} (GS)

\end{itemize}

%\pagebreak

\textbf{\em Science Goal \#2: Constrain the compositional evolution and transport rates within the rings.} The rates at which the rings' composition and large-scale structure evolve depend upon several parameters that are still not well constrained. The compositional evolution of the ring depends upon the flux and composition of the micrometeoroids that are constantly striking the rings, as well as the impact yield and the flow rates of different materials out of the rings in the form of vapor and fine dust grains. We also need to know what the non-icy components of the ring particles actually are, so that we can translate these input and loss rates into appropriate timescales for the rings. At the same time, while the large-scale structure of the rings does not change much over timescales of years, transport of material and angular momentum should cause the locations of select ring features to slowly change over time, and constraining this structural evolution can provide further information about the history and future of the ring system. Thus, we have the following measurement objectives:
\begin{itemize}
\vspace{-0.2cm} \item \textbf{Objective 2-1:} {\em Measure impact flux into rings and properties of impact ejecta clouds.} (IMG, MS, DD)
\vspace{-0.3cm}\item \textbf{Objective 2-2:} {\em Constrain temporal and spatial variability, and average value of the mass leaving and entering the rings.} (MS, DD)
\vspace{-0.3cm} \item \textbf{Objective 2-3:} {\em Measure water vapor fraction produced from impacts relative to ejected mass.} (MS)
\vspace{-0.3cm} \item \textbf{Objective 2-4:} {\em Constrain composition of non-icy ring material for comparison to primitive sources such as cometary/asteroidal impactors, and evolved sources such as organics at Titan and Enceladus.} (MS, DD)
\vspace{-0.3cm} \item \textbf{Objective 2-5:} {\em Observe secular radial evolution of ring material.} (IMG)

\end{itemize}

\textbf{\em Science Goal \#3: Document the interactions between the rings, magnetosphere and atmosphere.}
Many interactions between Saturn's atmosphere, magnetosphere, and rings involve plasma, neutral atoms, and dust being transported from the rings via the magnetosphere to the planet. Quantifying these interactions therefore requires measuring both the average densities of these materials, as well as how they vary with time and space. At the same time, electromagnetic forces and currents couple Saturn's upper atmosphere to the rings via the magnetic field. This not only influences how material flows between the planet and the rings, but also enables asymmetries in the magnetosphere to affect the structure of dusty material around the rings. In particular, both rotational and seasonal changes in the magnetospheric environment appear to influence the intensity of dusty features known as spokes over Saturn's B ring. Properly characterizing these connections requires in-situ measurements within the relevant environments to determine what aspects of the environment are stable and which change as the planet rotates. The measurement objectives associated with this goal are therefore:
 \begin{itemize}
\vspace{-0.2cm}
\item \textbf{Objective 3-1:} {\em Characterize the steady-state atmosphere of Saturn's rings.} (MS) 
\vspace{-0.3cm}
 \item \textbf{Objective 3-2:} {\em Determine the mechanisms and consequences of material transport between the rings and the upper atmosphere of Saturn.} (MS, DD)
\vspace{-0.3cm} \item \textbf{Objective 3-3:} {\em Determine the nature of electromagnetic coupling between Saturn and the rings.} (PW, MAG) 
\vspace{-0.3cm} \item \textbf{Objective 3-4:} {\em Determine the connection between properties of Saturn's magnetosphere, including its periodicities and composition, and dynamics in the rings.} (PW, MAG, IMG)
\vspace{-0.3cm} \item \textbf{Objective 3-5:} {\em Measure the properties and environment of spokes at close range and \textit{in~situ}.} (DD, PW, MAG, IMG)
\end{itemize}

\textbf{\em Science Goal \#4: Determine the relative importance of various dynamical processes in Saturn's atmosphere.}   Cloud morphology and composition can reveal much about basic processes like condensation, precipitation, upwelling, and photochemical haze production. At the same time, cloud motions provide information about the role of convection in transporting heat from below, aggregation of convective cells into giant storms, and connections with the planet-encircling jet streams.  They can also help identify the means by which turbulent dissipation is maintained and the role it plays in atmospheric circulation. Finally, knowing the altitude of features in the images allows one to distinguish deep convection from shallow cumulus and haze. Thus, we have the following measurement objectives:
\begin{itemize}
\vspace{-0.2cm} \item \textbf{Objective 4-1:} {\em Characterize the spatial and temporal distribution of $1-10$~km structures in Saturn's atmosphere.} (IMG)
\vspace{-0.3cm} \item \textbf{Objective 4-2:} {\em Characterize the processes that transport energy from the interior to the level where it is radiated to space.} (IMG)
\end{itemize}

\textbf{\em Science Goal  \#5: Characterize the variable and asymmetric aspects of Saturn's deep interior.}  
The locations and sizes of the structures responsible for the asymmetries in Saturn's gravity field are still unknown because Cassini was only able to detect these anomalies a handful of times. Mapping the planet's gravitational field repeatedly would allow structures with different rotation periods to be distinguished and quantified. In addition, Saturn's rings are now known to be sensitive to oscillations and asymmetries inside the planet, and so correlating relevant  ring structures with anomalies in the gravitational field will further constrain the magnitudes and lifetimes of asymmetric and time-variable structures in Saturn's interior. These considerations yield the following measurement objectives:
\begin{itemize}
\vspace{-0.2cm} \item \textbf{Objective 5-1:} {\em Characterize the longitudinal asymmetries and temporal variations in Saturn's gravity field.} (GS)
\vspace{-0.3cm} \item \textbf{Objective 5-2:} {\em Survey and monitor waves in Saturn's rings due to resonances with interior oscillation modes and mass anomalies.} (IMG)
\end{itemize}

\vspace{0.1cm}
{\bf Relationship to other Saturn system mission concepts:} In addition to being a scientifically valuable concept on its own, it is possible for a Ring Skimmer to be productively combined with other types of mission concepts to the Saturn system. Specifically, integrating a Saturn probe could be a good fit. A Saturn Ring Skimmer would naturally approach within 15,000~km of Saturn's equatorial cloud tops.  Higher latitudes could be reached at the beginning of the mission, while the spacecraft maneuvers its orbit into ring-skimming configuration. Furthermore, there is much synergy between the science goals of a probe and a skimmer, with a shared focus on Saturn's atmosphere in the context of its origin, interior, and the rest of the inner Saturn system. 

A Saturn Ring Skimmer could also conceivably observe Titan, Enceladus, or other moons.  Indeed, equatorial flybys of Titan are a required part of ring skimming, and close equatorial passes to Enceladus occur naturally and could be further targeted.  However, synergy in instrumentation and in science goals may be difficult to attain.  Moreover, flying over the scientifically interesting poles of the moons introduces vertical velocities that are not consistent with a ring-skimming orbit, which means that a mission would need to operate in phases in order to meet both moon-focused objectives and the objectives described in this paper, focusing alternately on either ring skimming or polar moon flybys but not both in the same mission phase.

\vspace{0.2cm}

{\bf Conclusions:} Our preliminary analysis demonstrates that a Ring-Skimmer mission has great scientific potential because it is able to address a wide range of high-level science goals that are relevant not only to Saturn and its rings, but also to giant planets and astrophysical disks in general. The basic mission concept therefore deserves further study to obtain robust cost estimates and to identify possible synergies with other potential future missions to the Saturn system. At the moment, we expect that a robust multi-disciplinary Ring Skimmer mission is best suited for the New Frontiers program, so we also ask that the Planetary Science Decadal Survey enable missions focused on the inner Saturn system to compete in future opportunities associated with that program.

%\vspace{0.8cm}
%\noindent {\bf References} 
%\bibliographystyle{aasjournal}
%\bibliography{bibliography}

\end{document}